\documentclass[aps,prl,twocolumn,groupedaddress]{revtex4}
\usepackage{graphicx,latexsym}

\begin{document}

\def\be{\begin{equation}}
\def\ee{\end{equation}}
\def\ba{\begin{eqnarray}}
\def\ea{\end{eqnarray}}

\title{Non-interacting electrons and the metal-insulator transition in 2D
with correlated impurities}
\author{M. Hilke}
\affiliation{Dpt. of Physics, McGill University, Montr\'eal,
Canada H3A 2T8}
\date{June 9, 2003}

\begin{abstract}
While standard scaling arguments show that a system of
non-interacting electrons in two dimensions and in the presence of
uncorrelated disorder is insulating, in this work we discuss the
case where inter-impurity correlations are included. We find that
for point-like impurities and an infinite inter-impurity
correlation length a mobility edge exists in 2D even if the
individual impurity potentials are random. In the uncorrelated
system we recover the scaling results, while in the intermediate
regime for length scales comparable to the correlation length, the
system behaves like a metal but with increasing fluctuations,
before strong localization eventually takes over for length scales
much larger than the correlation length. In the intermediate
regime, the relevant length scale is not given by the elastic
scattering length but by the inter-impurity correlation length,
with important consequences for high mobility systems.
\end{abstract}


\maketitle

It is generally believed that a non-interacting two dimensional
(2D) system in the presence of disorder is always insulating. This
result is based on extensive work on scaling theory pioneered by
the 'gang of four' \cite{abrahams}. It has also received
considerable amount of numerical support \cite{mckinnon}. The
general statement can be summarized as follows. In scaling theory
the localization length in 2D is given by $\L_c\simeq\lambda
e^{\pi k_F\lambda/2}$ \cite{lee}, where $\lambda$ is the elastic
scattering length. Hence, as soon as $\lambda$ is finite, the
localization length is finite. However, because $\lambda$ appears
in the exponent, when $k_F\lambda>>1$ the localization length can
be extremely large and difficult to probe experimentally. In the
standard Born approximation $\lambda^{-1}\sim n_I$, where $n_I$ is
the two-dimensional impurity concentration, hence, a non-zero
$n_I$ would lead to a finite $L_c$. At zero temperature a finite
localization length implies that the resistance diverges
exponentially for a system size exceeding the localization length.
Equivalently, for an infinite system, the resistance diverges when
the temperature, $T$ tends to zero because the phase coherence
length $l_\phi$ is infinite at zero $T$.

When considering discrete models based on Anderson's disordered
tight binding model \cite{lee2} very similar results are obtained.
All states are localized for any strength of disorder. These
results apply only if all sites are uncorrelated. Indeed, even in
1D there exist special long and short range correlations in the
disorder, which can lead to the existence of extended states in
these systems \cite{flores}. Similarly special systems can also be
found in 2D \cite{hilke}. The exact conditions under which these
localization conditions apply has recently gained considerable
interest because 2D electronic systems confined in a variety of
semiconducting structures, such as Si-MOSFETs and p-type
GaAs/AlGaAs have shown a strong metallic-like $T$-dependence of
the resistance \cite{kravchenko,hanein,simmons} down to the lowest
experimental $T$. In metallic, we understand a positive or
vanishing derivative of the resistivity as a function of the
temperature, i.e., $\partial\rho/\partial T\geq 0$.

Because in these systems disorder is always present, which is
expected to lead to localization, most of the recent theories have
considered interactions between electrons \cite{abrahams2} as a
possible mechanism for the metallic behavior. In this work, we
consider instead the case of correlations between impurities,
which leads to a genuine metal-insulator transition for an
infinite inter-impurity correlation length $l\rightarrow \infty$
for the non-interacting 2D disordered system (with random
potentials). Moreover, for a finite $l$, the crossover to
localization is determined by $l$ and not $\lambda$.

We start by considering the standard disordered tight binding
Anderson model in 2D:

\be
\psi_{n+1,m}+\psi_{n-1,m}+\psi_{n,m+1}+\psi_{n,m-1}=(E-V_{n,m})\psi_{n,m}.
\ee

When $V_{n,m}=0$ we recover the free particle case with solutions
of the type $\psi_{n,m}=e^{ikn}e^{ipm}$ and eigenvalues
$E=2\cos(k)+2\cos(p)$. When $V_{n,m}$ is random and uncorrelated,
scaling theory predicts that all states are localized for any
disorder strength. The same result holds in 1D. However, when the
potential is correlated the situation becomes very different. In
1D, when only every 2nd site is random, i.e., $V_{2n}=0$ and
$V_{2n+1}$ is random $\psi_{n}=\cos(n\pi/2)$ is an extended
solution of the disordered system, with energy $E=0$. In this
model only one energy has an infinite localization length and
since at all other energies all states are localized there is no
band of extended states \cite{hilke3}. However, when we consider a
similar potential in 2D something very interesting happens.

When defining $V_{2n,m}=0$ and $V_{2n+1,m}$ as random, we can
write a solution to equ. 1 as $\psi_{n,m}=\cos(n\pi/2)e^{ikm}$,
corresponding to $E=2\cos(k)$. Clearly, this is a fully extended
state. The difference in the 2D case compared to 1D is that we now
have a band of extended states for $-2\leq E\leq 2$. For every
energy in this band there is exactly one wave-function, which can
be written in this form, hence there is no degeneracy. Therefore,
when evaluating the two-terminal conductance of the system one
obtains $G=2e^2/h$ for $-2\leq E_{F}\leq 2$ and $G=0$ otherwise,
since all other states are localized. This gives rise to a band of
extended states around the band center very similar to the 3D
Anderson model, which leads to a metal-insulator transition when
the Fermi energy crosses $E=-2$ and $E=2$. Similarly, when only
the $V_{an,m}$'s are random, but all other potentials are zero,
the system has conductance steps at $E=\pm 2+2\cos(l\pi/a)$, with
$l=1,2,3,\ldots,a-1$ and $a$ an integer $a\geq 2$. We have
evaluated the conductance numerically as a function of the Fermi
energy and the results are presented in fig. 1. In order to
calculate the temperature dependence of the resistance we simply
used $G(T)=\int dEf'_T(E-E_F)G(E)$, where $f_T$ is the Fermi-Dirac
distribution function. For non-interacting electrons this is the
dominant temperature dependence since inelastic scattering is
strongly suppressed at low temperatures. Moreover, because $E_F$
is equivalent to the density, this system exhibits a genuine
metal-insulator transition as a function of density.

\begin{figure}
\vspace*{1.0cm}
\includegraphics[scale=0.25]{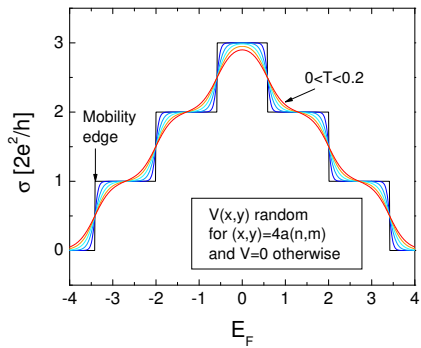}
\vspace*{-2cm} \caption{Fermi energy dependence of the 2-terminal
conductance with $V_{4n,m}$ random and all others zero. The
results are obtained from the tight-binding Hamiltonian and for
different temperatures.}
\end{figure}

The continuous case can be treated in an analogous way by
considering $\delta$-impurities with random amplitudes on a
lattice, which is similar to the 1D case discussed in ref.
\cite{hilke2}. In this case the random potential is
$V(x,y)=\sum_{n,m}V_{n,m}(y)\delta (na-x)\delta (ma-y)$, where
$V_{n,m}$ are random. The special energies for the conductance
steps are now given by $E=n^2$ in units of $\hbar^2\pi^2/2ma^2$,
where $a$ is the corresponding lattice constant of the random
impurities and $n=1,2,\cdots$. The corresponding conducting
wave-functions (extended states) are simply given by
$\psi(x,y)=sin(n\pi x/a)e^{iky}$ for $E=n^2+(ka/\pi)^2$. In the
inset of fig. 2 the zero temperature conductance, $G$, is shown as
a function of the Fermi energy.

\vspace*{-.0cm}
\begin{figure}
\includegraphics[scale=0.30]{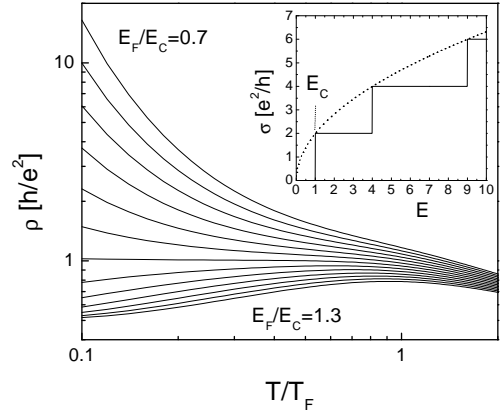}
\caption{Temperature dependence of the 2-terminal resistance for
various values of $E_F$. The inset shows the zero temperature
dependence of the conductance as a function of $E_F$ and the
dotted line illustrates the overall square root dependence of the
conductance steps}
\end{figure}

The temperature dependence is shown for different values of $E_F$
in fig. 2. Hence, in this simple model in which $\delta$
impurities are on a lattice with random amplitudes, we have a
$E_F$ induced metal-insulator transition at $E_C$, independent of
the disorder strength. The temperature dependence was obtained by
using the Fermi-Dirac distribution $R^{-1}=\int
dEf'_T(E-E_F)G(E)$. It is quite striking to note that the overall
shapes of the curves are very similar to the experimental ones in
high mobility 2D systems \cite{hanein}. The temperature scale in
these experiments is typically close to the Fermi temperature
$T_F$, which is of the order of 1K and very similar to the
relevant scale in our simple model.

In the above discussion we considered the two-terminal conductance
of the system. Experimentally, this is the quantity which is
normally measured. However, in order to neglect the contact
resistance, four terminal measurements are typically performed,
but at zero magnetic field this is equivalent to a two terminal
measurement assuming zero contact resistance. In addition, in 2D
the value of the conductance is converted to conductivity by
taking into account the geometry of the sample. Hence the
conductivity in our case would be $\sigma=G/\Box$, where $\Box$ is
the number of squares or geometrical factor. Since the conductance
of our model does not depend on the geometry or length of the
sample, our system would have an infinite conductivity if the
sample was infinitely long but finite in width and of zero
conductivity if infinitely wide and of finite length. At first
this does not seem physical but if one considers that each
experimental system has a finite $l_\phi$ at non-zero $T$, which
is equal in every direction for an isotropic system, the relevant
geometry is close to a square. This is very reasonable because for
any practical comparison with experiments, quantum effects such as
localization can only occur within $l_\phi$. Indeed, we could
model the experimental system as a network of quantum coherent
squares coupled classically and we would recover $\sigma=G$, which
is the expected result in which the value of the critical
conductivity does not depend on the geometry. This normalization
was used for all figures.

Experimentally, a large applied parallel magnetic field tends to
exponentially increase the resistivity \cite{yoon}. In a
non-interacting picture a parallel field would simply shift the
two spin subbands and therefore not significantly alter the
overall behavior, although it could decrease the critical
transition density. However, when comparing to experimental
systems in a parallel field other effects become very important
too. For instance, the 2D plane is not entirely flat because there
are micron sized in-plane deviations, which are of the order of 4
to 30nm high vertically \cite{willet}, which can allow random
fluxes to penetrate the 2D system and induces additional
localization. These effects are beyond the scope of this work and
will not be considered further.

Coming back to our model, where the disorder is confined to a
regular lattice, we could show that there is a well defined
metal-insulator transition, very similar to the experimental
situation. However, impurities do not usually form a perfect
crystal during the growth process. They are more likely to be
somewhat correlated but not to the extent to form a lattice. In
order to generalize our model, we start by considering the case,
where the impurities are uncorrelated $l=0$ and where the density
of impurities is low before we turn to the more general case of
arbitrary $l$.

For the numerical calculation we considered the tight-binding
equation (1) with finite extent $L$ in the $n$ direction and
infinite in $m$. Without disorder we have plane wave solutions of
the type $\psi^{(p)}_{n,m}=\sin(np\pi/L)e^{ikm}$, with
$E=2\cos(p\pi/L)+2\cos(k)$. This system can be viewed as a
quasi-1D quantum wire, where the degeneracy of states depends on
$L$ and $E$ and the conductance is simply $2e^2/h$ multiplied by
the degeneracy. In the middle of this system we now add $N_I$
punctual impurities in a rectangle of size
(width=L)$\times$(length=L). Since contacts are made by diffusing
a metal into the 2D layer, the metal side of the contact has a
much larger electron density than the 2D. Therefore, to model the
effect of contacts we have considered the following situation.
Instead of using a source and drain region of the same size as the
disordered region, we consider a much larger (essentially
infinite) contact point to the disordered region. The model is
illustrated in fig. 3, where the source and drain region has no
disorder and is much wider than the disordered region. Hence, the
wide regions correspond to the metal contacts, whereas the
disordered narrow region corresponds to the 2D system under study.
We evaluated numerically the conductance of the system using
standard transfer matrix techniques and show the conductance as a
function of the number of impurities for different sizes of the
disordered region in fig. 3. In our model we used a square
geometry $L\times L$, which has the particularity that at the band
center the conductance $G$ is independent of $L$ for $V=0$ and
$\sigma=G$. The curves were obtained by averaging over ten
different disorder configurations.

\begin{figure}
\vspace*{0cm}
\includegraphics[scale=0.25]{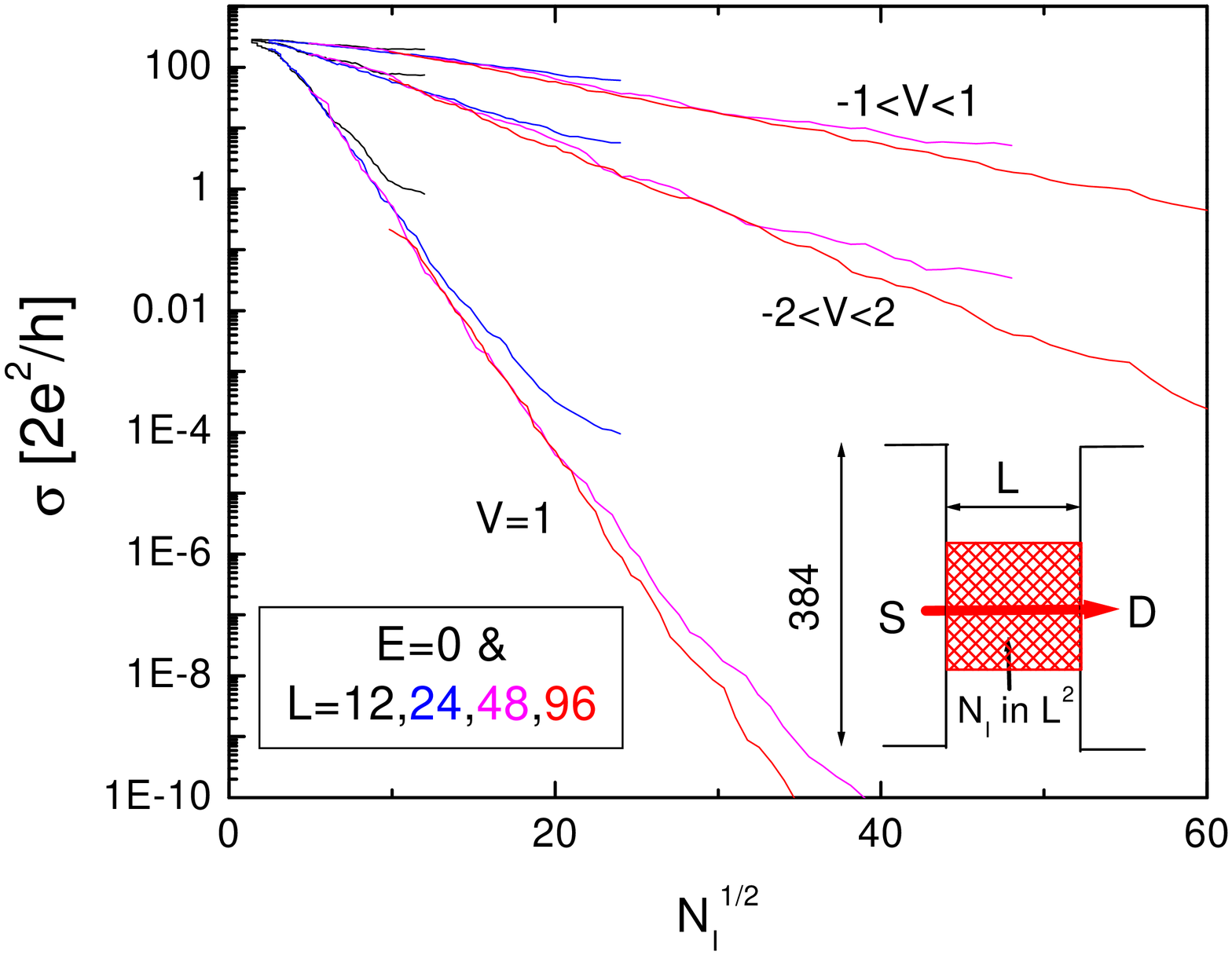}
\vspace*{-0cm} \caption{The conductivity as a function of the
number of impurities for different sizes and values of the
impurity potential, with $-0.5<V_1<0.5$, $-1<V_2<1$ and $V_3=1$. }
\end{figure}

Fig. 3, clearly shows that $\sigma$ only depends on the number of
impurities and not on the size of the system. Hence the dependence
is simply $\sigma(N_I)=\sigma(L^2n_I)$. This dependence holds for
the entire range until $\sigma$ saturates to the value where
$n_I\simeq 1$. At large enough number of impurities and in the
strong localization regime we have an excellent fit to the
expression $\sigma(N_I)\sim e^{-\alpha \sqrt{N_I}}\sim
e^{-\alpha\sqrt{n_I}L}$, where $\alpha$ depends only on the
impurity strength and $L_c^{-1}=\alpha\sqrt{n_I}$. The quality of
the fit and the simplicity of the fitting expression is quite
remarkable. This result also implies that for any non-zero
uncorrelated impurity density the system is always insulating,
which is consistent with scaling analysis.

However, when the impurities are distributed on a lattice, i.e.,
$l=\infty$, like in fig. 1, we also obtain a universal dependence
only on $N_I$ independent of $L$, i.e., $\sigma(N_I)$. The result
is shown in fig. 4. The only difference here is that
$\sigma(N_I\rightarrow \infty)=2e^2/h$ instead of zero. For this
data set we considered that only every second site is random,
which in this case leads to a metallic conductance of $2e^2/h$ at
the band center. This data set was obtained without averaging over
different disorder configurations and each data point represents
an independent impurity configuration.

\begin{figure}
\vspace*{2cm}
\includegraphics[scale=0.25]{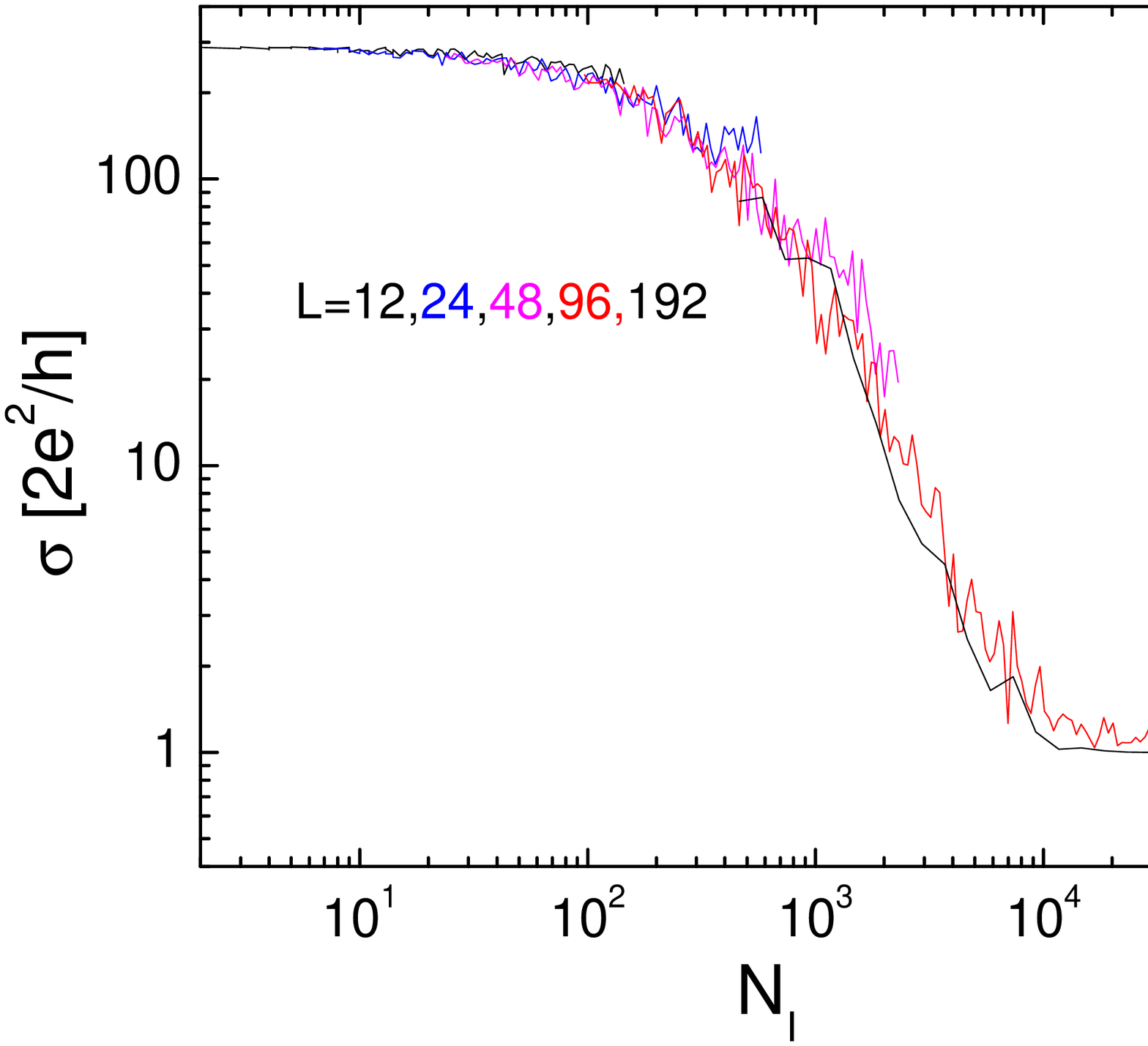}
\vspace*{-2cm} \caption{The conductivity as a function of the
number of impurities for different sizes and for $-1<V_2<1$. Here
only every second site is disordered.}
\end{figure}

We now consider the most interesting case, where we change the
inter-impurity correlation length $l$. The results in fig. 5
illustrate three cases ($l=0, l\neq 0, l=\infty$) for a fixed
system size $L$ since $\sigma$ does not depend explicitly on $L$
but only on $N_I$. In order to vary $l$ we chose to introduce two
types of impurities $N_I^{corr}$ and $N_I^{uncorr}$ and
$N_I=N_I^{corr}+N_I^{uncorr}$. The $N_I^{corr}$ impurities will
fall randomly on the lattice (of spacing 4 for fig. 5), whereas
the $N_I^{uncorr}$ impurities are not correlated to any site. The
correlation length is therefore given by the density of
uncorrelated impurities $l=L^2/N_I^{uncorr}$.

\begin{figure}
\vspace*{2cm}
\includegraphics[scale=0.25]{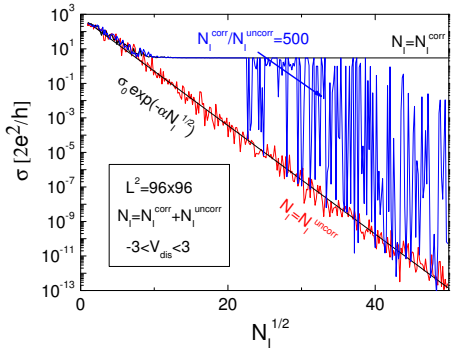}
\vspace*{-2cm} \caption{The conductivity as a function of the
number of impurities for different ratios of correlated to
uncorrelated impurities. The straight line is the fit $\sigma\sim
exp(-\alpha N_I^{1/2})$.}
\end{figure}

Overall, the conductivity remains metallic for length scales
comparable to the correlation length and the conductance
fluctuations are close to zero in this metallic regime. For length
scales exceeding the correlation length we first have increasing
conductance fluctuations before the system eventually localizes at
large enough length scales $L\gg l$ with a localization length
equal to the uncorrelated case. The fluctuations in the
intermediate regime are much larger than for the uncorrelated
system and constitute an important signature of the correlation in
the "metallic"-like state. Interestingly, a very similar
enhancement of the conductance fluctuations is seen close to the
transition from the metallic to insulating behavior in Si-MOSFET
systems and GaAs \cite{popovic}. In our model of correlated
impurities it is therefore the inter-impurity correlation length
$l$, which controls the range of metallic behavior and not the
semiclassical elastic scattering length $\lambda$, which can be
much smaller than $l$. Hence, a large $l$ can significantly
enhance the range of metallic behavior in 2D, which constitutes
one of the the main results of this article. This is in stark
contrast to the case where the impurity range is increased.
Indeed, in this case it was shown that a larger impurity range
enhances localization if impurities remain uncorrelated
\cite{beal}. The situation with correlated impurities was not
considered. Hence, this shows that inter-impurity correlations
suppress localization whereas large impurity ranges enhance
localization. In an applied perpendicular magnetic field, it was
recently shown that a long-range disorder potential might induce a
metal-insulator transition \cite{efetov} too. Clearly,
correlations between impurities have a significant impact on the
localization properties and have to be considered when comparing
to experimental systems.

In conclusion, we have analyzed the impact of inter-impurity
correlations on the localization properties of a non-interacting
2D electronic system. For impurities on a lattice and with random
potentials, corresponding to an infinite inter-impurity
correlation length, the system exhibits a true metal-insulator
transition. A finite correlation length enhances the scale over
which the system is conducting before the system is eventually
localized. A large correlation length can explain the large
conductance fluctuations and metallic behavior of experimental
systems. We also showed that the conductivity is only a function
of the number of impurities, $N_I$ and that the localization
length is given by $L_c\sim N_I^{-1/2}$ for $L\gg l$.

We would like to acknowledge helpful discussions with Boris Ischi
and Hong Guo and financial support from NSERC and FCAR.


\begin{thebibliography}{99}

\bibitem{abrahams} E. Abrahams, P.W. Anderson, D.C. Licciardello
and T.V. Ramakrishnan, {\em Phys. Rev. Lett.}, {\bf 42}, 673
(1979).
\bibitem{mckinnon} A. MacKinnon and B. Kramer, {\em Rep. Prog.
Phys.}, {\bf 56}, 1469 (1993).
\bibitem{lee} P.A. Lee and T.V. Ramakrishnan, {\em Rev. of Mod.
Phys.}, {\bf 57}, 287 (1985).
\bibitem{lee2} P.A. Lee and D.S. Fisher, {\em Phys. Rev. Lett.}, {\bf 47},
882 (1981).
\bibitem{flores}J.C. Flores, {\em J. Phys. Condens. Matter} {\bf 1}, 8471
(1989) and D.H. Dunlap, H.-L. Wu, and P. Phillips, {\em Phys. Rev.
Lett.} {\bf 65}, 88 (1990).
\bibitem{hilke} M. Hilke, {\em J. Phys. A: Math. Gen.} {\bf 27}, 4773
(1994)
\bibitem{kravchenko} S.V. Kravchenko {\em et al., Phys. Rev. B}
{\bf 50}, 8039 (1994).
\bibitem{hanein} Y. Hanein {\em et al., Phys. Rev. Lett.} {\bf
80}, 1288 (1998).
\bibitem{simmons} M.Y. Simmons {\em et al., Phys. Rev. Lett.} {\bf
80}, 1292 (1998).
\bibitem{abrahams2} E. Abrahams, S. V. Kravchenko, M. P. Sarachik,
{\em Rev. Mod. Phys.} 73, 251 (2001).
\bibitem{hilke3} M. Hilke, {\em J. Phys. A: Math. Gen.}, {\bf 30}, L367(1997).
\bibitem{hilke2} M. Hilke and J.C. Flores {\em Phys. Rev. B} {\bf 55}, 10625
(1997).
\bibitem{yoon} J. Yoon {\em et al., Phys. Rev. Lett.} {\bf
82}, 1744 (1999).
\bibitem{willet} R. L. Willett, J.W.P. Hsu, D. Natelson, K.W. West, and L.N.
Pfeiffer {\em Phys. Rev. Lett.} {\bf 87}, 126803 (2001).
\bibitem{popovic} S. Bogdanovich and D. Popovic, {\em Phys. Rev. Lett.}
{\bf 88}, 236401 (2002) and R. Leturcq {\em et al., Phys. Rev.
Lett.} {\bf 88}, 076402 (2003). 
\bibitem{beal} M.T. B\'eal-Monod, A. Theumann and G. Forgacs {\em Phys. Rev. B} {\bf 46},
15726 (1992).
\bibitem{efetov} D. Taras-Semchuk1 and K.B. Efetov {\em Phys. Rev. B} {\bf 64}, 115301
(2001). 




\end{thebibliography}
\end{document}